\documentclass[prb,twocolumn,showpacs]{revtex4}
\usepackage{graphicx}
\usepackage{amsmath}
\usepackage{amssymb}
\usepackage{bbold}

\DeclareMathAlphabet{\bi}{OML}{cmm}{b}{it}
\begin{document}


\title{Thermoelectric probe for Rashba spin-orbit interaction strength 
in a two dimensional electron gas}

\author{SK Firoz Islam and Tarun Kanti Ghosh}
\affiliation{Department of Physics, Indian Institute of Technology-Kanpur,
Kanpur-208 016, India}

\begin{abstract}
Thermoelectric coefficients of a two dimensional electron gas (2DEG) with 
the Rashba spin-orbit interaction (SOI) are presented here. 
In absence of magnetic field, thermoelectric coefficients are enhanced 
due to the Rashba SOI.
In presence of magnetic field, the thermoelectric coefficients of 
spin-up and spin-down electrons oscillate with different frequency
and produces beating patterns in the components of the total 
thermoelectric power and the total thermal conductivity. 
We also provide analytical expressions of the thermoelectric 
coefficients to explain the beating pattern formation. 
We obtain a simple relation which determines the Rashba SOI strength if
the magnetic fields corresponding to any two successive beat nodes are 
known from the experiment.
\end{abstract}


\date{\today}
\maketitle

\section{Introduction}
There has been a rapid growth of research interest on the Rashba 
SOI in low-dimensional condensed matter system 
after the proposal of spin field effect transistor by Datta and Das 
\cite{das1}.  
This is due to the possible applications in spintronics devices
\cite{appl1,appl2,appl3}.
The SOI is responsible for other interesting effects like 
spin Hall effect \cite{she}, spin dynamics and zitterbewegung 
\cite{zb,spin,zb1}.  
In the narrow gap semiconductor heterostructures, the dominant Rashba 
SOI \cite{rashba,rashba1} appears due to the asymmetric quantum wells. 
The Rashba SOI strength is proportional to the internally generated crystal field.
This strength can also be enhanced by applying suitable electric field 
perpendicular to the plane of the electron's motion \cite{tech,matsu}. 

A pseudo Zeeman effect occurs at finite momentum of the electron
due to the Rashba SOI even in absence of magnetic field.
A direct manifestation of pseudo Zeeman effect due to the SOI
is a regular beating pattern in the magnetoelectric transport
measurements such as Shubnikov-de Hass (SdH) oscillations \cite{beat_exp}
in 2DEG. These oscillations occur due to
two closely spaced different frequency of spin-up and spin-down
electrons. The Rashba SOI strength is determined by analyzing the
beating patterns in the SdH oscillations 
\cite{miller,datta1}.
The SOI was determined by fitting the experimental data with the
model calculations for the SdH oscillations. Later, many realistic
approach was considered and the estimated strength is in good agreement
with the extrapolated results \cite{alter1,alter2,firoz}. Recently, there is 
an interesting proposal \cite{firoz1} of determining the Rashba SOI 
strength by analyzing the beating patterns in the Weiss oscillations 
\cite{weiss,weiss1}.

On the other hand, thermoelectric properties of materials \cite{nolas} 
have attracted considerable interest from both experimental and 
theoretical point of view due to potential applications in technology
\cite{application,application1}.
There is a strong effect of perpendicular magnetic field on thermal
transport properties of any system. Therefore, the magnetothermal 
coefficients can be used as an additional probe.
In presence of perpendicular magnetic field, the diffusing charge carriers 
experience the Lorentz force. This produces a transverse electric
field in addition to the longitudinal electric field. The longitudinal
thermopower or the Seebeck coefficient is defined as
$S_{xx} = - \frac{\nabla V_x}{\nabla T}$. On the other hand, the
transverse thermopower or the Nernst coefficient is defined as
$ S_{xy} = - \frac{\nabla V_x}{\nabla T}$. Here,
$ \nabla V_x$ and $\nabla V_y $ are the induced voltage generated
by the thermal gradient and the magnetic field, respectively.
Theoretical and experimental studies on thermoelectric coefficients of 
2DEG systems in presence of magnetic field started after the discovery 
of the quantum Hall effect. 
In most of the thermoelectric measurements of 2DEG systems, 
the thermopower is being measured since the 
thermal resistivity of a 2DEG is extremely high.
The Nernst coefficient is quite sensitive to various properties of the
systems e.g. shape of the Fermi surface as well as electron mean free path
\cite{behnia}. It is being used as a probe to study various strongly correlated
electron systems such as Kondo lattices 
\cite{behnia1} and graphene field effect transistors \cite{graphene,graphene1}.
Moreover, the thermopower $S$ and the thermal conductivity $\kappa$ are used 
as the metrics to measure the thermoelectric performance \cite{behnia}.
In addition to these, we will show here that magntethermoelectric coefficients 
can also be used to determine the Rashba SOI strength.

There are mainly two mechanisms contribute to the thermal conductivity and
the thermopower, namely the thermodiffusion and phonon drag.
Generally, the phonon drag contribution is vanishingly small at very
low temperature.
In absence of the magnetic field, the diffusive thermopower has been 
continuously reported in the low range of temperature
\cite{kundu,exp1,syme,rafael,reno,epl,seebek}.
In presence of magnetic field, the oscillation of the diffusive 
thermopower has been studied theoretically as well as
experimentally \cite{prb86,prb95,topical,maximov,arindam}.
It is seen in the low magnetic field regime that both 
$S_{xx} $ and $S_{xy}$ are periodic in inverse of the magnetic field.
This is due to the oscillating density of states of the 2DEG in presence
of magnetic field.

There is no theoretical or experimental study on magnetothermoelectric
properties of the 2DEG systems with the Rashba SOI.
We report here for the first time the effect of the Rashba SOI on 
thermal transport properties of a 2DEG in presence of perpendicular magnetic field.
The total thermalconductivity and the total thermopower produce beating
patterns because the thermoelectric coefficients for spin-up and
spin-down electrons oscillate with two closely spaced different frequencies.
By analyzing the beating pattern, we find a simple equation which determines
the Rashba SOI strength if the magnetic fields corresponding to any two successive 
beat nodes and the number of oscillations in between are known from the experiment.

This paper is organized as follows. In section II, we briefly mention
the energy spectrum and the DOS of the 2DEG with the Rashba SOI for
zero and non-zero magnetic field cases.  In section III, we have studied
the thermoelectric coefficients for zero magnetic field case. We also 
provide the formalism to be used for studying thermoelectric coefficients
in presence of magnetic field. 
In section IV, we present our numerical and analytical results.
We provide a summary and conclusion of our work in section V.

\section{ENERGY SPECTRUM AND DENSITY OF STATES of a 2DEG with the Rashba SOI}

\subsection{Zero magnetic field case}
The Hamiltonian of an electron with the Rashba SOI 
is given by \cite{rashba}
\begin{equation}
H = \frac{{\bf p}^2}{2m^{\ast }} \mathbb{1}   + 
\frac{\alpha}{\hbar }({\mbox{\boldmath $\sigma$} }  \times p)_z, 
\end{equation}
where ${\bf p}$ is the two-dimensional momentum operator, 
$m^{\ast }$ is the effective mass of the electron, 
$ \mathbb{1}$ is the unit matrix,  
${\mbox{\boldmath $\sigma$} }=(\sigma_x,\sigma_y,\sigma_z)$ 
are the Pauli spin matrices and $\alpha $ is the strength of 
the Rashba SOI. At non-zero momentum, the spin degeneracy is lifted 
due to the presence of the SOI.
The energy spectrum of the ''spin-up" and ''spin-down" electron is 
given by 
\begin{equation}
E^{\pm} = \frac{\hbar^2k^2}{2m^*} \pm \alpha \mid k \mid.
\end{equation}
Here, the + and - signs correspond to the spin-up and spin-down electrons.
The density of states (DOS) \cite{winkler} for spin-up and spin-down electrons 
are
\begin{equation} \label{dos1}
g^{+}(E) = \frac{D_0}{2} 
\Big[1+\sqrt{\frac{E_{\alpha}}{E_{\alpha}+4E}}\Big] \Theta(E)
\end{equation}
 
and

\begin{eqnarray} \label{dos2}
g^{-}(E) & =& \frac{D_0}{2} 
\Big[1-\sqrt{\frac{E_{\alpha}}{E_{\alpha}+4E}}\Big] \Theta(E) 
\nonumber \\ & + &
D_0\sqrt{\frac{E_{\alpha}}{E_{\alpha}+4E}} 
\Theta(-E) \Theta(E+E_{\alpha}/4).
\end{eqnarray}
Here, $D_0=m^*/(\pi\hbar^2)$,
$ E_{\alpha} = 2 m^{\ast} \alpha^2/\hbar^2$ is the Rashba energy 
determined by the Rashba SOI strength $\alpha $ and $\Theta(E) $ 
is the unit step function.

\subsection{Non-zero magnetic field case}
The Hamiltonian of an electron $(-e)$ with the Rashba SOI 
in presence of a perpendicular magnetic field ${\bf B} = B \hat z$ is given by
\begin{equation} \label{Ham}
H = \frac{({\bf p} + e {\bf A})^2}{2m^{\ast }} \mathbb{1} + 
\frac{\alpha}{\hbar }\left[ {\mbox{\boldmath $\sigma$} } \times 
({\bf p}+e{\bf A})\right]_z + \frac{1}{2}g\mu _{_B} B \sigma_z, 
\end{equation}
where $ \mu_{_B} = e \hbar/(2m_e) $ is the Bohr magneton with 
$m_e$ is the free electron mass and $g$ is the effective Lande 
$g$-factor. 
The exact energy spectrum and the corresponding eigenfunctions of the
above Hamiltonian are derived in Ref. \cite{alter1}.
The resulting eigenstates are labeled by a new quantum number $s$.
For $s=0$, there is only one energy level which is same as the lowest Landau
level without the Rashba SOI. The corresponding energy is given by
$ E_0^+ = E_0 = (\hbar \omega - g \mu_{_B} B)/2 $.
Here, $ \omega = eB/m^* $ is cyclotron frequency.
For $s=1,2,3....$, there are two branches of the energy levels, 
denoted by $+$ corresponding to the "spin-up" electrons 
and $-$ corresponding to the "spin-down" electrons with energies
\begin{equation}
E_s^{\pm} = s\hbar\omega{\pm} \sqrt{E_0^2 + s E_{\alpha} \hbar \omega}.
\end{equation}

Using the Green's function method, the DOS for spin-up and spin-down 
electrons in presence of magnetic field are calculated in Ref. \cite{firoz}.
These are given by
\begin{eqnarray} \label{dos3}
D^{\pm}(E) & \backsimeq & \frac{D_0}{2} 
\Big[1 + 2 \exp{\Big\{-2\Big(\frac{\pi\Gamma_0}{\hbar\omega}\Big)^2\Big\}} 
\nonumber \\ 
& \times & \cos{\Big\{\frac{2\pi}{\hbar\omega}
\Big(E+\frac{E_{\alpha}}{2}\mp \sqrt{E_0^2+E_{\alpha}E}\Big)\Big\}}\Big],
\end{eqnarray}
where $\Gamma_0 $ is the impurity induced Landau level broadening.

\section{Thermoelectric coefficients}
In this section, we shall develop the formalism for the thermoelectric 
coefficients of a 2DEG with the Rashba SOI system for both the cases:
zero and non-zero magnetic fields.

\subsection{Zero magnetic field case}
In this sub-section, we consider a 2DEG with the Rashba SOI and calculate
the thermal power and thermal conductivity.
Within the linear response regime, the electrical current 
density ${\bf J} $ and the thermal current density ${\bf J}_{q}$ 
for spin-up and spin-down electrons can be written as
\begin{equation}
{\bf J}_{\pm} = L_{\pm}^{11} {\bf E} + L_{\pm}^{12} (-\nabla T)
\end{equation}
and
\begin{equation}
{\bf J}_{\pm}^{q} = L_{\pm}^{21} {\bf E} + L_{\pm}^{22}(-\nabla T),
\end{equation}
where ${\bf E} $ is the electric field and $ L_{\pm}^{ij} $ with $ i,j=1,2$ are 
the phenomenological transport coefficients for spin-up and 
spin-down electrons in absence of magnetic field. 
These are the main equations that determine the response of a system to 
the external forces such as electric field and temperature gradient.
In presence of the Rashba SOI, the spin-up and spin-down electrons will 
contribute to the total electrical and thermal current. Therefore, 
the total electrical current and the thermal current densities are
\begin{equation}
{\bf J} = L^{11} {\bf E} + L^{12}(-\nabla T)
\end{equation}
and
\begin{equation}
{\bf J}^{q}=L^{21} {\bf E} + L^{22}(-\nabla T).
\end{equation}
Here, $L^{ij} = L_{+}^{ij} + L_{-}^{ij}$
and $L^{ij}$ can be written in terms of the integral
$I^{(r)}$: $L^{11}=I^{(0)}, L^{21} = TL^{12} = -I^{(1)}/e$, 
$L^{22}=I^{(2)}/(e^2T)$. Also, $I^{(r)}=I^{(r),+}+I^{(r),-}$ with
\begin{equation}
I^{(r),\pm} = \int dE  \Big[-\frac{\partial f(E)}{\partial E}\Big]
(E-\eta)^{r}  \sigma^{\pm}(E),
\end{equation}
where $r=0,1,2$ and $f(E)=1/[1+exp(E-\eta)\beta]$ is the 
Fermi-Dirac distribution function with $\eta$ is the chemical 
potential and  $\beta=1/(k_{_B}T)$. Here, 
$\sigma^+(E) $ and $ \sigma^{-}(E)$ are the energy-dependent conductivity
for spin-up and spin-down electrons, respectively. 
In an open circuit condition ($J=0$),
the thermopower is given by $S=L^{12}/L^{11}$.
Then at low temperature, diffusion thermopower $S$  and the  
diffusion thermal conductivity $\kappa$ can be expressed in
terms of the electrical conductivity through the Mott's relation 
and the Wiedemann-Franz law as
\begin{equation} \label{tp}
S = - L_0 e T \Big[ \frac{d}{dE}\ln\sigma(E) \Big]_{_{E=E_F}}
\end{equation}
and
\begin{equation}
\kappa = L_0 T \sigma(E_{_F}).
\end{equation}
Here, $ L_0 = (\pi^2 k_{_B}^2)/(3e^2) $ is the Lorentz
number and $\sigma(E_{_F}) = \sigma^+ (E_{_F}) + \sigma^- (E_{_F})$ 
is the total electrical conductivity at the Fermi energy.

By using the Boltzmann transport equation, we 
evaluate the zero-temperature energy-dependent electrical conductivity 
for spin-up and spin-down electrons, which are given by 
\begin{equation} \label{con}
\sigma^{\pm}(E)= \frac{e^2}{m^*} \tau(E)g^{\pm}(E)
\Big[ E + \frac{E_{\alpha}}{4} \Big].
\end{equation}
Assuming the energy dependent scattering time to be 
$\tau=\tau_0 (E/E_{_F})^{p}$, where $p$ is a constant depending on the
scattering mechanism. We also assumed that $\tau $ is the same for spin-up and
spin-down electrons.
Substituting Eqs. (\ref{dos1}), (\ref{dos2}) and (\ref{con}) into Eq. (\ref{tp}),   
then the diffusion thermopower is obtained as   
\begin{equation}
S= - L_0 \frac{e T}{E_F} \Big[p+1 - \frac{E_{\alpha}}{4E_F}\Big].
\end{equation}

We calculate the total electrical conductivity $\sigma (E_{_F})$ at 
the Fermi level, which is given as 
\begin{equation}
\sigma (E_{_F})= \frac{n e^2\tau_0}{m^*} +  \frac{m^*e^2 \tau_0 \alpha^2}{2\pi \hbar^4}
= \sigma_0\Big[ 1 + \frac{E_{\alpha}}{4E_F}\Big],
\end{equation}
where $ \sigma_0 = \frac{n e^2\tau_0}{m^*} $ is the Drude conductivity without SOI. 
The similar expression of the Drude conductivity is obtained by using a different method
in Ref. \cite{vasi}.
The total thermal conductivity is then
\begin{equation}
\kappa = L_0 T \sigma_0\Big( 1 +  \frac{E_{\alpha}}{4E_F}\Big).
\end{equation}
We note that the thermal conductivity and the thermopower is enhanced 
due to the presence of the Rashba SOI.

\subsection{Non-zero magnetic field case}

In this subsection, we shall study the thermoelectric coefficients
of a 2DEG with the Rashba SOI in presence of the perpendicular magnetic field.
Thermoelectric coefficients in presence of magnetic field (without SOI)
were obtained by modifying the Kubo formula in Ref. \cite{streda,oji}. 
Here we shall
generalize these results to the SOI systems.
These phenomenological transport coefficients can be re-written as
\begin{equation}
\sigma_{\mu\nu}^{\pm} = {\cal L}_{\mu\nu}^{(0),\pm}
\end{equation}
\begin{equation}
S_{\mu\nu}^{\pm} = \frac{1}{eT}[({\cal L}^{(0),\pm})^{-1}{\cal L}^{(1),\pm}]_{\mu\nu}
\end{equation}
\begin{equation}\label{cond1}
\kappa_{\mu\nu}^{\pm} = \frac{1}{e^2T}[{\cal L}_{\mu\nu}^{(2),\pm}- 
eT({\cal L}^{(1),\pm}S^{\pm})_{\mu\nu}],
\end{equation}
where
\begin{equation}\label{Lmunu}
{\cal L}_{\mu\nu}^{(r),\pm}=\int dE  \Big[-\frac{\partial f(E)}{\partial E}\Big]
(E-\eta)^{r}  \sigma_{\mu\nu}^{\pm}(E).
\end{equation}
Here, $\mu,\nu=x,y$. Also, $\sigma_{\mu \nu}^{\pm}(E)$,
$S_{\mu \nu}^{\pm} $ and $\kappa_{\mu \nu}^{\pm}$
are the zero-temperature energy-dependent conductivity, thermopower and
thermal conductivity tensors, respectively, for spin-up and spin-down electrons.
The total thermopower and thermal conductivity can be obtained from 
$S_{\mu \nu} = S_{\mu \nu}^{+}  + S_{\mu \nu}^{-}$ and 
$\kappa_{\mu \nu} =  \kappa_{\mu \nu}^{+} + \kappa_{\mu \nu}^{-}$.

In electron systems, conduction of carriers takes place
by the diffusive and collisional mechanisms. The collisional 
contribution leads to the SdH oscillation with inverse magnetic 
field due to the quantized nature of the energy spectrum. 
We will consider the collisional mechanism only because
electrons do not possess any drift velocity in our case.
In the linear response regime,
the conductivity tensor can be written as the sum of 
diagonal and non-diagonal as 
$\sigma_{\mu \nu} = \sigma_{\mu \nu}^{\rm d} + \sigma_{\mu \nu}^{{\rm nd}}$, 
where $\sigma_{\mu\nu}^{\rm nd}$ is the Hall
contribution. Here, $\sigma_{xx}=\sigma_{xx}^{\rm col}$ and
$\sigma_{yy}=\sigma_{xx}^{\rm col}+\sigma_{yy}^{\rm dif} 
= \sigma_{xx}^{\rm col}$. 
Similarly, for the thermal transport coefficients 
the following relations are valid: 
$ {\cal L}_{xx}^{(r)}={\cal L}_{xx}^{(r){\rm col}} = 
{\cal L}_{yy}^{(r){\rm col}}$ and
$ {\cal L}_{yy}^{(r)}={\cal L}_{yy}^{(r){\rm dif}} + 
{\cal L}_{yy}^{(r){\rm col}}={\cal L}_{yy}^{(r){\rm col}}$.
The exact form of the finite temperature collisional conductivity 
has been calculated in Ref. \cite{alter1} for the 
screened impurity potential 
$ U({\bf q}) = 2 \pi e^2/(\epsilon \sqrt{q_{x}^2 + q_{y}^2 + k_s^2})$ 
in momentum space.
Here, $k_s$ is the inverse screening length and $ \epsilon $ is 
the dielectric constant of the material. 
In the limit of small $|{\bf q}| \ll k_s $, 
$ U({\bf q}) \simeq 2\pi e^2/(\epsilon k_s) = U_0$. In this limit, one can 
use $\tau_0^2 \approx \pi l^2\hbar^2/N_IU_0^2$ with
$\tau_0$ is the collisional time, $l=\sqrt{\hbar/eB}$ is the magnetic 
length scale,
$U_0$ is the strength of the screened impurity potential and 
$N_I$ is the two-dimensional impurity density.
The exact form of the finite temperature conductivity 
can be reduced to the zero-temperature energy-dependent electrical conductivity as
\begin{equation} \label{exact}
\sigma_{xx}^{\pm}(E) = \frac{e^2}{h} 
\frac{ N_I U_0^2}{2\pi \Gamma_0 l^2}I_{s}^{\pm},
\end{equation}
where $I_s^{\pm} = [(2s\mp1)D_s^4-2sD_s^2+(2s\pm1)]/A_s^2 $ with
$ D_s = \sqrt{s E_{\alpha} \hbar \omega}/[E_0 + \sqrt{E_0^2 + s E_{\alpha} 
\hbar \omega}]$ and $A_s = 1 + D_s^2 $.
Using Eq. (\ref{Lmunu}), the finite temperature diagonal and off-diagonal 
coefficients (${\cal L}_{xx}^{(r)} $ and  $ {\cal L}_{yx}^{(r)}$) can be written as
\begin{equation}
{\cal L}_{xx}^{(r),\pm} = 
\frac{e^2}{h}\frac{ N_I U_0^2}{2\pi \Gamma_0 l^2}
\sum_{s}I_{s}^{\pm}\Big[(E-\eta)^{r}
\Big(-\frac{\partial f(E)}{\partial E}\Big)\Big]_{E=E_s^{\pm}} 
\end{equation}
and
\begin{eqnarray}
{\cal L}_{yx}^{(r),\pm} & = & \frac{e^2}{h}
\sum_{s} 
\frac{\Big[D_{s+1}(D_s\sqrt{s} \mp \sqrt{2} k_{\alpha}l) + 
\sqrt{s+1}\Big]^2}{{A_sA_{s+1}}}  \nonumber \\ 
& \times &
\int_{E_s^{\pm}}^{E_{s+1}^{\pm}}dE\Big[(E-\eta)^{r}
\Big(-\frac{\partial f(E)}{\partial E}\Big)\Big]_{E=E_s^{\pm}}.
\end{eqnarray}

\section{Numerical results and discussions}

In our numerical calculations, the following parameters are used:
carrier concentration $n_e=3 \times 10^{15}$ /m${}^2$, 
effective mass $m^{*}=0.05 m_e$ with $m_e$ is the free electron mass,  
$g = 4 $ and the Rashba SOI strength $\alpha = 5 \times 10^{-12} $ eV-m and
$\Gamma_0=0.01$ meV. 
For better visualization of the oscillations, 
we have used $T=1$ K for the thermopower, $T=0.5$ K for the thermal conductivity.   
In Fig. [1], the components of the thermopower tensor in units of $-k_{_B}/e$ are shown as
a function of the inverse magnetic field. The diagonal thermopower components 
$S_{xx} $ and $S_{yy}$ are identical and therefore only $S_{xx}$ is shown.
In Fig. [2], the thermal conductivity is shown
as a function of the inverse magnetic field. 
The magnetic field dependence of the
thermal conductivity is same as that of the electrical conductivity.
Figures [1] and [2] show the appearance of the beating pattern in the thermopower
and thermal conductivity.

\begin{figure}[t]
\begin{center}\leavevmode
\includegraphics[width=98mm]{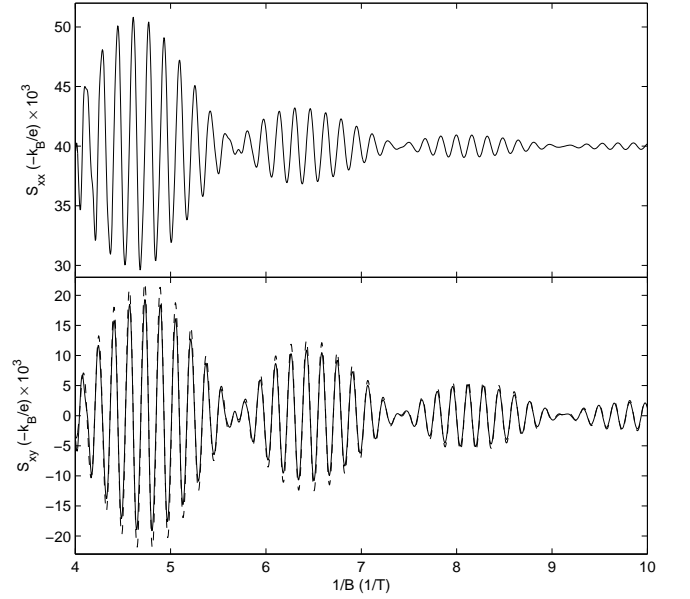}
\caption{Plots of the thermopower versus inverse magnetic field.
In the lower panel, dashed and solid lines correspond to the
analytical and exact results, respectively.}
\label{Fig1}
\end{center}
\end{figure}

\begin{figure}[t]
\begin{center}\leavevmode
\includegraphics[width=98mm]{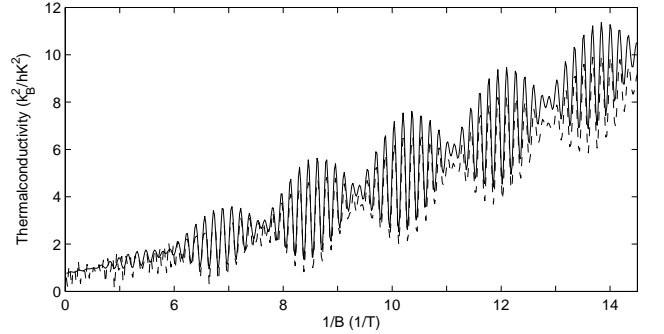}
\caption{Plots of the longitudinal component of thermalconductiviy 
$\kappa_{xx} $ versus inverse magnetic field $B$. 
The  dashed and solid lines correspond to the exact numerical and analytical results.}
\label{Fig1}
\end{center}
\end{figure}

To analyze the beating pattern in the thermoelectric coefficients,
we shall derive analytical expressions of the thermoelectric coefficients.
The components of the thermopower for spin-up and spin-down electrons are given by
\begin{equation} \label{Syy}
S_{xx}^{\pm} = S_{yy}^{\pm} = \frac{1}{eT}
\Big[ \frac{\sigma_{xx}^{\pm}}{S_0} {\cal L}_{xx}^{(1),\pm}+
\frac{{\cal L}_{yx}^{(1),\pm}}{\sigma_{yx}^{\pm}}\Big]
\end{equation}
and
\begin{equation} \label{Sxy}
S_{xy}^{\pm} = - S_{yx}^{\pm} = \frac{1}{eT}
\Big[\frac{\sigma_{xx}^{\pm}}{S_0} (-{\cal L}_{xy}^{(1),\pm})+
\frac{{\cal L}_{xx}^{(1),\pm}}{\sigma_{yx}^{\pm}} \Big].
\end{equation}
The dominating term in the above two equations is the last term. 
The analytical form of $\kappa_{xx}$ and $S_{xy}$ can be obtained directly by 
deriving analytical form of the phenomenological transport coefficients. 
The analytical form of the DOS given in Eq. (\ref{dos3}) allows us to obtain 
asymptotic expressions of $ S_{xy} $ and $ \kappa_{xx}$.
This is done by replacing the summation over discrete quantum 
numbers $s$ by the integration i.e; 
$ \sum_s \rightarrow 2\pi l^2 \int D^{\pm}(E)dE $,
then we get
\begin{equation}
{\cal L}_{xx}^{(1),\pm} \simeq \Big(\frac{-\pi}{\beta}\Big)
\frac{\sigma_0}{8(\omega\tau_0)^2}\Omega_{D} G^{\prime}(x)
\sin\Big(2\pi\frac{f^{\pm}}{B}\Big)
\end{equation}
and
\begin{equation} \label{cond2}
{\cal L}_{xx}^{(2),\pm} \simeq 
\Big(\frac{\pi}{\beta}\Big)^2
\frac{\sigma_0}{8(\omega\tau_0)^2}\Big[\frac{1}{3} - 
\frac{\Omega_{D}}{2} G^{\prime \prime} (x) 
\cos\Big(2\pi\frac{f^{\pm}}{B}\Big)\Big],
\end{equation}
where the impurity induced damping factor is
\begin{equation}
\Omega_{D}=2\exp\Big\{-2\Big(\frac{\pi\Gamma_0}{\hbar\omega}\Big)^2\Big\}
\end{equation}
and the temperature dependent damping factor is the derivative of 
the function $G(x) $ with $ G(x)=x/\sinh(x)$.
Here, $x=T/T_c$ with $T_c=\hbar\omega/2\pi^2k_{_B}$.
Note that $G(x)$ is the temperature dependent damping factor for the electrical 
conductivity tensor.
Also, the oscillation frequencies are
\begin{equation}
f^{\pm} = \frac{m^*}{\hbar e} \Big[E_{F} +
\frac{E_{\alpha}}{2} \mp \sqrt{E_0^2 + E_{\alpha} E_{F}} \Big].
\end{equation}

The off-diagonal thermopower $S_{xy} $ for spin-up and spin-down electron is 
obtained as
\begin{equation} \label{offd_s}
S_{xy}^{\pm} = - \frac{k_{_B}}{e} 
\frac{\pi}{4\omega\tau_0}\Omega_D G^{\prime}(x) 
\sin\Big(2\pi\frac{f^{\pm}}{B}\Big).
\end{equation}

The total thermopower is given as
\begin{equation}\label{ana_s}
S_{xy} = -\frac{ k_{_B}}{e}  
\frac{\pi}{2\omega\tau_0} \Omega_D G^{\prime} (x) 
\sin\Big(2\pi\frac{f_{\rm a}}{B}\Big)
\cos\Big(2\pi\frac{f_{\rm d}}{B}\Big).
\end{equation}
Here, $f_{\rm a} = (f^{+}+f^{-})/2$ and $f_{\rm d} =(f^{+}-f^{-})/2$.
In the lower panel of Fig. [1], we compare the analytical expression of
$S_{xy} $ with that of the numerical result. 
The analytical result matches very well with the numerical results.

For thermal conductivity, the dominant term in 
$\kappa_{xx}$  is $ {\cal L}_{xx}^{(2)}$. 
The approximate analytical form of $ \kappa_{xx}^{\pm}$ can be 
obtained from Eqs. (\ref{cond1}) and (\ref{cond2}) as
\begin{equation} \label{diagok}
\kappa_{xx}^{\pm} \simeq L_0 \frac{T \sigma_0}{8(\omega \tau_0)^2}
\Big[1 - \frac{3}{2} \Omega_D G^{\prime \prime}(x) 
\cos\Big(2\pi\frac{f^{\pm}}{B}\Big)\Big].
\end{equation}
The total thermal conductivity can be written as
\begin{eqnarray}\label{ana_k}
\kappa_{xx} & \simeq &  L_0 
\frac{\sigma_0 T}{4(\omega \tau_0)^2}
\Big[1- \frac{3}{2} \Omega_D G^{\prime \prime}(x) \nonumber \\
& \times & 
\cos\Big(2\pi\frac{f_{\rm a}}{B}\Big)\cos\Big(2\pi\frac{f_{\rm d}}{B}\Big)\Big].
\end{eqnarray}

Equations (\ref{offd_s}) and (\ref{diagok}) show that the thermopower and 
the thermalconductivity of spin-up and spin-down electron oscillates 
with different frequency $f^+$ and $f^{-}$, respectively.
Therefore, the beating pattern appears in the total $S_{xy}$ and $\kappa $.
It is quite difficult to obtain the analytical expression of $S_{yy}$, 
but the origin of the oscillatory part is due to the oscillatory density 
of states at the Fermi energy.

We get the condition for beating nodes from the periodic term with
frequency difference  $f_{\rm d}$: $\cos(2\pi f_{\rm d}/B)_{B=B_j}=0$ which gives
\begin{equation} \label{alpha}
\sqrt{\Delta_s^2 + (1-g^*)^2 (\hbar \omega_j)^2} = \hbar\omega_j(j+\frac{1}{2}).
\end{equation}
Here, $  \Delta_s = 2k_{_F}\alpha $ is the zero-field 
spin splitting energy with $k_{_F}$ is the Fermi wave vector,
$j=1,2,3..$ is the j-th beat node and
$ g^* = g m^*/(2m_e)$. Also,
$\omega_j = e B_j/m^*$ and $B_j$ is the magnetic field corresponding to the
$j$-th beat node. 
Using the above equation, one can determine the zero-field 
spin splitting energy or the Rashba strength if we know the number ($j$) of any
node and the corresponding magnetic field $B_j$.
In practice, the numbering of the beat nodes is quite difficult.
The above equation can be re-written for two successive beating nodes as 
\begin{equation} \label{main}
\sqrt{ \Big(\frac{\Delta_s}{\hbar \omega_{j+1}}\Big)^2 + ( 1 - g^*)^2}
- \sqrt{\Big(\frac{\Delta_s}{\hbar \omega_{j}}\Big)^2 + ( 1 - g^*)^2}
= 1.
\end{equation}
Therefore, the Rashba SOI strength can be determined from  Eq. (\ref{main})
by knowing the magnetic fields correspond to any two successive beat nodes.

In the above analytical expressions [Eqs. (\ref{ana_s}) and (\ref{ana_k})]   
the periodic term with 
frequency $f_{\rm a}$ gives the number of oscillations between the 
two successive beat nodes as given by 
\begin{equation} \label{num_osc}
N_{\rm osc} = \frac{m^{*}}{e\hbar}
\Big(E_F+\frac{E_{\alpha}}{2}\Big)\Big(\frac{1}{B_{j+1}}
-\frac{1}{B_{j}}\Big).
\end{equation}
Therefore, we can also determine the Rashba strength from Eq. (\ref{num_osc}) 
by knowing the magnetic fields correspond to any two successive beat 
nodes and the number of oscillations in between.
We note that Eqs. (\ref{alpha}) and (\ref{num_osc}) are the same as
obtained in the beating pattern formation in the SdH oscillations
\cite{firoz}.

\section{conclusion}
We present theoretical study of the effect of the Rashba SOI on the 
thermoelectric coefficients.
In absence of magnetic field, the thermopower and the thermal conductivity
are enhanced due to the presence of the SOI.
The numerical results of all the thermoelectric coefficients are given.
In addition to the numerical results, we provide the analytical expressions 
of the off-diagonal component of the thermopower $(S_{xy})$ and the diagonal 
components of the thermal conductivity ($\kappa_{xx}$). 
The appearance of the 
beating pattern in the thermoelectric coefficients can be explained from the 
fact that the two branches
oscillate with slightly different frequency and produce beating pattern in the
thermoelectric coefficients.
The analytical results match very well with the numerical results.
The Rashba SOI strength can be determined if the magnetic field corresponding to
any two successive beat nodes are known from the experiment.

\section{Acknowledgement}
This work is financially supported by the CSIR, Govt.of India under the grant
CSIR-SRF-09/092(0687) 2009/EMR F-O746.


\begin{thebibliography}{55}


\bibitem{das1}
S. Datta and B. Das,
Appl. Phys. Lett. {\bf 56}, 665 (1990)

\bibitem{appl1}
I. Zutic, J. Fabian, and S. Das Sarma,
Rev. Mod. Phys. {\bf 76}, 323 (2004)

\bibitem{appl2}
A. Wolf et , 
Science {\bf 294}, 1488 (2002)

\bibitem{appl3}
D. D. Awschalom and M. E. Flatte, 
Nature Phys {\bf 3}, 153 (2007)

\bibitem{she}
S. Murakami, N. Nagaosa, and S. C. Zhang,
Science {\bf 301}, 1348 (2003)


\bibitem{zb}
J. Schliemann, D. Loss, and R. M. Westervelt,
Phys. Rev. Lett. {\bf 94}, 206801 (2005)

\bibitem{spin}
B. C. Hsu and J. S. V. Huele,
Phys. Rev. B {\bf 80}, 235309 (2009)

\bibitem{zb1}
T. Biswas and T. K. Ghosh, 
J. Phys.: Condens. Matter {\bf 24}, 185304 (2012)


\bibitem{rashba}
E. I. Rashba and V. I. Sheka,
Dokl. Akad. Nauk SSSR {\bf 2}, 162 (1959);
E. I. Rashba, Sov. Phys. Solid State {\bf 2}, 1109 (1960)


\bibitem{rashba1}
Y A Bychkov and E I Rashba,
J. Phys. C: Solid State, {\bf 17}, 580 (1984)


\bibitem{tech}
J. Nitta, T. Akazaki, H. Takayanagi, and T. Enoki,
Phys. Rev. Lett. {\bf 78}, 1335 (1997)

\bibitem{matsu}
T. Matsuyama, R. Kursten, C. Meibner, and U. Merkt,
Phys. Rev. B {\bf 61}, 15588 (2000)


\bibitem{beat_exp}
J. Luo, H. Munekata, F. F. Fang, and P. J. Stiles,
Phys. Rev. B {\bf 38}, 10142 (1988); {\bf 41}, 7685 (1990)

\bibitem{miller}
B. Das, D. C. Miller, S. Datta, R. Reifenberger, W. P. Hong,
P. K. Bhattacharya, J. Sing, and M. Jaffe,
Phys. Rev. B {\bf 39}, 1411 (1989)

\bibitem{datta1}
B. Das, S. Datta, and R. Reifenberger, 
Phys. Rev. B {\bf 41}, 8278 (1990)

\bibitem{alter1}
X. F. Wang and P. Vasilopoulos, 
Phys. Rev. B {\bf 67}, 085313 (2003)

\bibitem{alter2}
S. G. Novokshonov and A. G. Groshev,
Phys. Rev. B {\bf 74}, 245333 (2006)

\bibitem{firoz}
SK Firoz Islam and T. K. Ghosh,
J. Phys.: Condens. Matter {\bf24}, 035302 (2012)


\bibitem{firoz1}
SK Firoz Islam and T. K. Ghosh,
J. of Phys.: Condens. Matter {\bf 24}, 185303 (2012)

\bibitem{weiss}
D. Weiss, K. von Klitzing, K. Ploog, and G. Weimann,
Europhys. Lett. {\bf 8}, 179 (1989)  

\bibitem{weiss1}
F. M. Peeters and P. Vasilopoulos,
Phys. Rev. B {\bf 46}, 4667 (1992)



\bibitem{nolas}
G. S. Nolas, J. Sharp, and H. J. Goldsmid, 
{\it Thermoelectrics} (Springer-Verlag, Berlin, 2001)


\bibitem{application}
F. J. DiSalvo, 
Science {\bf 285}, 703 (1999)

\bibitem{application1}
G. J.  Snyder and E. S. Toberer,
Nature Mater {\bf 7}, 105 (2008)


\bibitem{behnia}
K. Behnia, M. -A. Measson, and Y. Kopelevich,
Phys. Rev. Lett. {\bf 98}, 076603 (2007)


\bibitem{behnia1}
R. Bel, K. Behnia, Y. Nakajima, K. Izawa, 
Y. Matsuda, H. Shishido, R. Settai, and Y. Onuki,
Phys. Rev. Lett. {\bf 92}, 217002 (2004)


\bibitem{graphene}
Y. M. Zuev, W. Chang, and P. Kim,
Phys. Rev. Lett. {\bf 102}, 096807 (2009)


\bibitem{graphene1}
P. Wei, W. Bao, Y. Pu, C. N. Lau, and J. Shi,
Phys. Rev. Lett. {\bf 102}, 166808 (2009)

\bibitem{kundu}
S. Kundu, C. K. Sarkar, and P. K. Basu,
J. Appl. Phys. {\bf 61}, 5080 (1987)


\bibitem{exp1}
R. T. Syme, M. J. Kellyt, and M. Pepper,
J. Phys.: Condens. Matter {\bf 1}, 3375 (1989)

\bibitem{syme}
R. T. Syme and M. J. Kearney,
Phys. Rev. B {\bf 46}, 7662 (1992)


\bibitem{rafael}
C. Rafael, R. Fletcher, P. T. Coleridge, Y. Feng, and Z. R. Wasilewski,
Semicond. Sci. Technol. {\bf 19}, 1291 (2004)

\bibitem{reno}
W. E. Chickering, J. P. Eisenstein, and J. L. Reno,
Phys. Rev. Lett. {\bf 103}, 046807 (2009)

\bibitem{epl}
A. Gold and V. T. Dolgopolov,
Europhys. Lett. {\bf 96}, 27007 (2011)

\bibitem{seebek}
S. Y. Liu, X. L. Lei, Norman, and J. M. Horing,
arXiv:1106.1262v1


\bibitem{prb86}
R. Fletcher, J. C. Maan, K. Ploog, and G. Weimann,
Phys. Rev. B {\bf 33}, 7122 (1986)

\bibitem{prb95}
R. Fletcher, P. T. Coleridge, and Y. Feng, 
Phys. Rev. B {\bf 52}, 2823 (1995) 

\bibitem{topical}
R. Fletcher, Semicond. Sci. Technol. {\bf 14}, R1 (1999) 


\bibitem{maximov}
S. Maximov, M. Gbordzoe, H. Buhmann, L. W. Molenkamp, and D. Reuter, 
Phys. Rev. B {\bf 70}, 121308 (R) (2004)

\bibitem{arindam}
S. Goswami, C. Siegert, M. Pepper, I. Farrer, D. A. Ritchie, and A. Ghosh,
Phys. Rev. B {\bf 83}, 073302 (2011)

\bibitem{winkler}
Spin-orbit coupling effects in two-dimensional electron and
hole systems by R. Winkler, Springer

\bibitem{vasi}
P. M. Krstajic, M. Pagano, and P. Vasilopoulos, 
Physica E, {\bf 43}, 893 (2011)

\bibitem{streda}
L. Smreka and P. Streda, 
J. Phys. C: Solid State Phys. {\bf 10}, 2153 (1977)

\bibitem{oji}
H. Oji, 
J. Phys. C: Solid State Phys., {\bf 17}, 3059 (1984)


\end{thebibliography}
\end{document}